\documentstyle[epsfig,twoside,fleqn,espcrc2]{article}

\newcommand{\AmS}{{\protect\the\textfont2
  A\kern-.1667em\lower.5ex\hbox{M}\kern-.125emS}}
\newcommand {\ignore}[1]{}
\def\app#1#2#3{          {\it Astropart. Phys. }{\bf #1} (19#2) #3}
\title{Neutrino electron scattering and left-right symmetry: 
	future tests}
\author{O. G. Miranda\thanks{On leave from Departamento de F\'{\i}sica 
	CINVESTAV-IPN}
     V. Semikoz \thanks{On sabbatical leave from Izmiran, Moscow}
     and Jos\'e W. F. Valle
	\thanks{http://neutrinos.uv.es}
\address{Instituto de F\'{\i}sica Corpuscular - C.S.I.C.\\
   Departament de F\'{\i}sica Te\`orica, Universitat de Val\`encia\\
   46100 Burjassot, Val\`encia, Spain}}

\begin{document}

\begin{abstract}
Low-energy high-resolution neutrino-electron scattering experiments
may play an important role in testing the gauge structure of the
electroweak interaction. We propose the use of radioactive neutrino
sources (e.g. $^{51}$Cr) in underground experiments such as BOREXINO and 
HELLAZ as a probe of the weak neutral current
structure. As an illustration, we display the sensitivity of these
detectors in testing the possible existence of right-handed weak
neutral currents.
\end{abstract}

\maketitle

Probing the gauge structure of the electroweak interaction using
accelerators or reactor neutrinos has already been common practice for
a number of years.  Accelerators provide mostly muon neutrino beams
and their precision in determining the neutral current parameters is
well below present LEP accuracy.  The first observation of $\nu_{e} e
\to \nu_{e} e$ scattering at LAMPF \cite{Allen1} resulted in a total
of $236 \pm 35$ events \cite{Allen2} leading to a total cross section
of $(10.0 \pm 1.5 \pm 0.9) \times 10^{-45} cm^2\times E_{\nu}(MeV)$
for a neutrino mean energy of 30 MeV.  This measurement showed that
the interference term between neutral and charged currents is
negative, as predicted by the standard model.  However, it is very
hard to achieve a precision measurement that is able to probe very
accurately the parameters of the weak neutral current. As for reactor
experiments, one is limited by the accuracy of electron 
anti-neutrino fluxes as well as by geometry.  

\ignore{ The $\nu_{e} e \to \nu_{e} e$ scattering process has proved
to be an useful tool in studying the $^8B$ neutrinos coming from the
sun at underground installations.  The electron recoil spectrum has
been measured in the Kamiokande Cerenkov detector with a threshold
energy of 7.5 MeV.  Superkamiokande should be able to reach a threshold
energy of 5 MeV and an energy resolution of about 20 \% .  Determining
the electron recoil spectrum should be also one of the goals to be
pursued at the future Sudbury Neutrino Observatory.  }

Here we suggest the possibility of studying $\nu_{e} e \to
\nu_{e} e$ scattering from terrestrial neutrino sources with improved
statistics.  A similar idea has been proposed as a test of
non-standard neutrino electro-magnetic properties, such as magnetic
moments \cite{Vogel}.  In contrast to reactor experiments, a small
radioactive isotope source can be surrounded by gas or liquid
scintillator detectors with full geometrical coverage.  Here we show
how a low-energy high-resolution experiment may play a role in testing
the structure of the neutral current weak interaction.  The
ingredients for doing such experiments are either already available
(e.~g. the chromium source has already been used for calibrating the
GALLEX experiment \cite{GALLEX}) or under investigation (NaI detectors
have already been used in dark matter searches \cite{Lama} and the
BOREXINO and HELLAZ \cite{Borexino,Hellaz} detectors have been
extensively discussed).  These detectors should reach good energy
resolution and relatively low threshold.  Some cryogenic detectors have
also been mentioned in this context \cite{cryo}.

\vskip .4cm

Here we explicitly determine the sensitivity of these radioactive
neutrino source experiments as probes of the gauge structure
of the electroweak interaction, with special emphasis on left-right
symmetric extensions of the standard electroweak theory.

We assume a generic electroweak gauge model in which the main
contributions to the neutrino electron scattering cross section arise
from the exchange of charged and neutral intermediate vector bosons,
i.e. from charged (CC) and neutral currents (NC).

The charged current amplitude for the $\nu_{e} e\to\nu_{e} e $  
process can be written, after a Fierz transformation as 
\begin{equation}
{\cal M_{CC}} = \sqrt{2}G_{F} \bar{\nu}\gamma^{\nu}(1-\gamma^5)\nu 
	\bar{e}\gamma_{\mu} c_{L}\frac{1-\gamma^5}{2} e. \label{CCA}
\end{equation}
For the specific case of the Standard Model (SM) we have $c_{L}=1$.

On the other hand, the neutral current contribution to the 
amplitude for the $\nu_{e} e\to\nu_{e} e $ process can be given as 
\begin{eqnarray}
{\cal M_{NC}} &=& \sqrt{2}G_{F} \bar{\nu}\gamma^{\mu}(1-\gamma^5) \nu
	\big\{ \bar{e}\gamma_{\mu}\big[ g_{L}\frac{1-\gamma^5}{2}+ \nonumber \\
	&+&  g_{R} \frac{1+\gamma^5}{2}\big]\big\} e. \label{NCA} 
\end{eqnarray}

In this case the tree level (SM) prediction is
$g_{L,R}=\frac12 (g_{V}\pm g_{A})$, $g_{V}=-1/2+2Sin^2\theta_{W}$ and
$g_{A}=-1/2$. The values of these coupling constants have been well
measured from $e^{+}e^{-} \to l^{+}l^{-}$ at high energies by the LEP
Collaborations.  A combined fit from LEP results at the Z peak gives
\cite{Stickland} $g_{V}=-0.03805\pm 0.00059$ and $g_{A}=-0.50098 \pm
0.00033$.  These results have given strong constraints on right-handed
neutral currents.  Radiative electroweak corrections for the SM at low
energies were recently computed \cite{Sirlin} and are included in
computations.

The $\nu_e e\to \nu_e e$ differential cross section following from
(\ref{CCA}) and (\ref{NCA}) in terms of $T/\omega_1$ is given by
\begin{eqnarray}
\frac{d\sigma}{dT} &=& \frac{2m_{e}G^2_{F}}{\pi}\big\{ 
	(g_{L}+c_{L})^2+ 
	g_{R}^2 (1-\frac{T}{\omega_1})^2 - \nonumber \\ 
	&-& (g_{L}+c_{L})g_{R}\frac{m_e}{\omega_1}
            \frac{T}{\omega_1} \big\}\label{DCS}
\end{eqnarray}

Here, $T$ is the recoil electron energy, and $\omega_1$ is the
neutrino energy; therefore $T/\omega_1 <1$ for any value of
$\omega_1$. It is also important to note that we have in this
expression the ratio $m_e/ \omega_1$, which is important for low
energies and negligible for accelerator neutrino energies.
All terms in this cross section are potentially sensitive to the
right-handed current admixtures.  As we will see neutrino-electron
scattering experiments with sufficiently low energies and high
resolution may provide a novel way to test for the presence of
right-handed neutral currents.

\vskip .4cm

We now turn to a brief discussion on the models based on the
Left-Right Symmetric gauge group $ SU (2)_L \otimes SU (2)_R \otimes U
(1)_{B - L} \cite{LR1,LR2}$.  These are attractive since they offer
the possibility of incorporating parity violation on the same footing
as gauge symmetry breaking, instead of by hand as in the SM.  The
Lagrangian for neutral currents in the left-right symmetric model (LRSM)
is given by
\begin{equation}
-{\cal L}= g J^{3^{\mu}}_{T_{L}} W_{L}^{3^{\mu}}
+ gJ^{3^{\mu}}_{T_{R}} W_{R}^{3^{\mu}}
+g' J^{\mu}_{{Y}} B^{3^{\mu}}
\end{equation}
where 
$J^{3^{\mu}}_{T_{L,R}}=\bar\psi_{L,R} \gamma^{\mu} T^3_{L,R} \psi_{L,R}$
and $T^3_{L,R}$ is the third isospin component. 

After applying the corresponding neutral gauge boson diagonalization
matrix to get the mass eigenstates we have
\begin{eqnarray}
{\cal L} &=&gs_W\big[ J_{\rm em}A- \frac1{c_W} (a_1J_{L}^Z+
         b_1J^Z_{R})Z_1 +\label{prd} \\
        &+&\frac1{c_W} (a_2J_L^Z+b_2J^Z_R)Z_2 \big]   \nonumber
\end{eqnarray}
where $J_{L,R}=J_{L,R}^3-\sin^2\theta_W J_{em}$ and 
\begin{eqnarray}
a_1=s_W \frac{s_{\phi}}{r_W} -\frac{c_{\phi}}{s_W}&,& 
\qquad b_1=s_ {\phi} \frac{c^2_W}{s_Wr_W} , \\ \nonumber
a_2=\frac{s_Wc_{\phi}}{r_W}+\frac{s_{\phi}}{s_W}&,& 
\qquad b_2= \frac{c_{\phi}c^2_W}{s_Wr_W} . \nonumber
\end{eqnarray}
Here $s_W$, $c_W$ and $r_W$ is a shorthand notation for
$sin\theta_W$, $cos\theta_W$ and $\sqrt{cos2\theta_W}$; 
and $s_\phi$, $c_\phi$ for $sin\phi$,
$cos\phi$ with $\phi$ the mixing angle between $Z_L$ and $Z_R$.

In the low energy limit, the $\nu_e e\to \nu_e e$ amplitude is given,
for left handed $\nu_e$ by
\begin{equation}
{\cal M}_{NC}^{LR} = \frac{8G_{F}}{\sqrt{2}} 
\big[AJ^{\nu}_{L}J^{e}_{L}+ BJ^{\nu}_{L}J^{e}_{R}  \big] \label{ALR}
\end{equation}
with 
\begin{eqnarray}
A &=& s^2_W(a^2_1+\gamma a^2_2) \nonumber \\
B &=& s^2_W(a_1b_1+\gamma a_2b_2) \nonumber 
\end{eqnarray}
and $\gamma =\big(\frac{M_{Z_1}}{M_{Z_2}} \big)^2 $.

Comparing the amplitude in Eq. (\ref{ALR}) with the one shown in Eq.
(\ref{NCA}) we can identify the expressions for the neutral current
constants in this model as
\begin{eqnarray}
g_{L}&=&A\frac12(g_{V}+g_{A})+B\frac12(g_{V}-g_{A}) \nonumber \\
g_{R}&=&A\frac12(g_{V}-g_{A})+B\frac12(g_{V}+g_{A}). \label{LRG}
\end{eqnarray}
The dependence on the LRSM parameters is contained in $A$ and
$B$. Therefore, the differential cross section in the LRSM will be
that of Eq (\ref{DCS}) with the values for $g_{L}$ and $g_{R}$ given
by Eq. (\ref{LRG}).

Note that $\nu_e e\to \nu_e e$ scattering is not sensitive to
right-handed charged currents because the interference term between
the corresponding amplitude and the SM one is suppressed either by the
neutrino mass (Dirac case) or by the mixing with the heavy neutrinos
(Majorana case, seesaw model). In fact, this this just an example of
the general situation that one finds when trying to constrain
charged-current parameters via purely leptonic processes (see
\cite{Holstein} and \cite{Wolfenstein}).

Neutral current couplings have been measured with a great accuracy at
the Z peak in LEP and they give strong bounds on the mixing angle
$\phi$ between massive neutral gauge bosons (the allowed range being
$- 0.015 < \phi < 0.002 $) \cite{LEP}. On the other hand the mass of
the additional gauge boson is constrained by direct searches or by
electroweak fits \cite{lang}. In the latter case the constraint on the
mass of the extra neutral gauge boson translates into $\gamma \leq
0.055$ at 95 \% C. L. \cite{lang}.  Equivalent constraints coming from
neutrino experiments were derived by the CHARM II Collaboration
\cite{charm} for $\nu_\mu e\to \nu_\mu e$ scattering, leading to
$\gamma \leq 0.130$ at 95 \% C. L., far worse than the one obtained in
a global electroweak fit.

\vskip .4cm

Here we want to show that new low-energy high-precision $\nu_e e\to
\nu_e e$ scattering experiments could improve this constraint. Two
ingredients are required for this kind of experiment: a strong
low-energy electron-neutrino source and a high-precision detector.
Here we focus on the potential of radioactive neutrino sources
(e.g. $^{51}$Cr) in underground detectors as a  probe of the
weak neutral current structure.  The first strong low-energy neutrino
source has been recently prepared for the calibration of the GALLEX
neutrino experiment \cite{GALLEX}. This was a $^{51}$Cr source with an
activity of $1.67 \pm 0.03$ MCi. The main neutrino lines in this
source correspond to 0.746 MeV (81 \% ) and 0.751 MeV (9 \%).

At present $\nu_{e} e \to \nu_{e} e$ scattering has not been measured
at energies below 1 MeV.  However, there are several proposals in this
direction. Here we concentrate in BOREXINO and HELLAZ.  These will be
sensitive to the desired range of electron recoil energy. The expected
energy thresholds are 250 KeV and 100 KeV respectively.  As for
the energy resolution, BOREXINO should reach 12 \% at 250 KeV and 8 \%
at 500 KeV, so we take bins of 50 KeV width. In the case of HELLAZ a
mean energy resolution of few percent is envisaged; therefore, we
consider 10 KeV bins.  Thus BOREXINO and HELLAZ can detect the two
main lines of the ${}^{51}$Cr. Here we estimate the expected
differential number of events both for the SM and the left-right
symmetric model, using the expression
\begin{eqnarray}
\frac{dN}{dT}& =& \Phi N_{e} \Delta t \big\{0.81 
          \left( \frac{d\sigma}{dT}  \right)_{E_{\nu}=746 KeV} + \nonumber \\
       &+& 0.09 \left( \frac{d\sigma}{dT}  \right)_{E_{\nu}=751 KeV}\big\}.
\end{eqnarray}
where we have assumed that these detectors surround the neutrino
source, as in the case of GALLEX \cite{GALLEX} with an exposure time
$\Delta t$ of twenty days.

In order to have an idea of the expected sensitivity on $\gamma$ as a
function of the hypothetical measurement, we must guess the
measurement numbers including their errors, as well as the energy
resolution attained in the detectors.

Let us assume that the detectors will measure the Standard Model
prediction with a hypothetical error $\sigma_{N_{i}}$ for bin $i$. In
this case we can determine the corresponding constraint on $\gamma$
as a function of $\sigma_{N_{i}}$. In Fig 1 we display the constraint
at 95 \% C. L. which can be attained on the ratio $\gamma $ as a
function of $\sigma_{N_{i}}$ (in percent).

Due to the high expected number of events in BOREXINO, the expected
statistical error for $N_{i}$ is of the order of
0.5 \%.  Therefore, the constraining power in this case is mostly
dependent on the systematic error. If the systematic error can be kept
at the level of 3 \%, then it will be possible to get a bound on
$\gamma$ of the same order as obtained by CHARM II for the case
of $\nu_\mu$ scattering. 

In contrast for the case of HELLAZ we expect to achieve a better
energy resolution, leading to a lower slope than the previous case, as
displayed in the figure. Thus in this case one correspondingly obtains
a better sensitivity to the parameter $\gamma$.

\begin{figure}[t]
%\begin{center}
\mbox{\epsfig{file=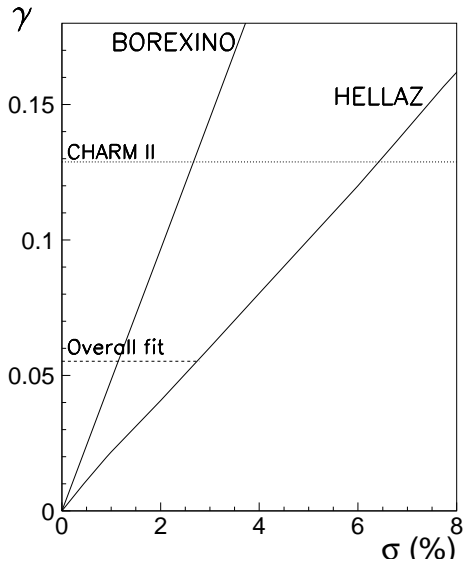,width=\linewidth}}
\\
%\vspace{0.5cm}
%Figure \ref{fig1}
%\end{center}
%\null
%\vspace{-1.5cm}
%\null
%\refstepcounter{figure}
\caption{Constraining power on $\gamma$ for different experiments depending 
on the experimental error}
%\label{fig1}
\end{figure}

Another possibility is the LAMA proposal \cite{Lama}.  This experiment
proposes to use a one ton NaI detector with similar characteristics to
the 100 Kg version already used for dark matter searches.  The
proposed source is an anti-neutrino ${}^{147}$Pm source that could
reach an activity of 5 MCi and a half-life of 2.6 years. In contrast
to the $^{51}$Cr this source has a continuous spectrum instead of
discrete lines.  The maximum neutrino energy will be 234.7 KeV.  In
this case one expects a better energy resolution than for the HELLAZ
case and, therefore, an enhanced sensitivity on the parameter $\gamma$
is expected. A detailed analysis of this case is now in progress
\cite{msv}.

\vskip 0.4cm

In summary, we conclude that a new generation of low-energy solar
neutrino-type detectors using strong radioactive neutrino sources may
open new experimental possibilities in testing the structure of the
electroweak interaction.

\vskip 0.4cm

We would like to thank useful discussions with Rita Bernabei, Miguel
Angel Garc\'{\i}a-Jare\~no, Concha Gonz\'alez-Garc\'{\i}a, Philippe
Gorodetzky, Francis von Feilitzsch, Sandra Malvezzi, Stanislav
Mikheev, Stephan Schonert, and Tom Ypsilantis.  This work was
supported by DGICYT under grant number PB95-1077 by the TMR network
grant ERBFMRXCT960090 of the European Union. O. G. M. was supported by
a CONACYT fellowship from the mexican government, and V. S. by the
sabbatical grant SAB95-506 and RFFR gronts 97-02-16501, 95-02-03724.


\begin{thebibliography}{9}

\bibitem{Allen1} 
Allen {\it et. al.}, Phys. Rev. Lett. 55  (1985) 2401.

\bibitem{Allen2}
 Allen {\it et. al.}, Phys. Rev. D47  (1993) 11.
 
\bibitem{Vogel}
Vogel and Engel,  Phys. Rev. D39  (1989) 3378. 

\bibitem{GALLEX} 
GALLEX Coll.,  Phys. Lett. B342 (1995) 440.

\bibitem{Lama}
R. Bernabei, these proceedings; R. Bernabei {\it et. al.}, Phys. Lett. B389 
(1996) 757; I. R. Barabanov {\it et. al.}, ROM2F-97/21, submitted to 
Astroparticle Physics.

\bibitem{Borexino}
J. B. Benziger {\it et. al.}, A proposal for participating in the Borexino 
solar neutrino experiment, October 30, 1996.

\bibitem{Hellaz} 
F. Arzarello {\it et. al.}: CERN-LAA/94-19.

\bibitem{cryo}
A. Alessandrello et al \app{3}{95}{239}

\bibitem{Stickland} 
D. Stickland, {Electroweak results from $e^{+}e^{-} $ colliders}, 
these proceedings.

\bibitem{Sirlin} 
John N. Bahcall, Marc Kamionkowski, Alberto Sirlin, 
    Phys. Rev. D51  (1995) 6146.

%\bibitem{331} 
%M. Singer, J. W. F. Valle, J. Schechter, 
%    Phys. Rev. D22  (1980) 738.
%
%J. W. F. Valle, M. Singer 
%    Phys. Rev. D28  (1983) 540.

\bibitem{LR1}
J.C. Pati and A. Salam, 
    Phys. Rev. D10  (1975) 275.

R.N. Mohapatra and J.C. Pati, 
    Phys. Rev. D11  (1975) 2558.


\bibitem{LR2}
R.N. Mohapatra and G. Senjanovi\'c, 
    Phys. Rev. D23  (1981) 165 and references therein. 

\bibitem{Holstein}
Barry R. Holstein,  Weak Interactions in Nuclei, Princeton series in 
physics, Princeton University Press, 1989.

\bibitem{Wolfenstein} 
L. Wolfenstein,
	 Phys. Rev. D29 (1984) 2130.


\bibitem{LEP}
O. Adriani {\it et. al.}, L3 Coll., Phys. Lett. B306 (1993) 187.

\bibitem{lang}
M. Cvetic, P. Langacker, hep-ph/9707451.

\bibitem{charm}
P. Vilain, {\it et. al.}, CHARM II Coll., Phys. Lett. B332 (1994) 465.

\bibitem{msv} O. G. Miranda, V. Semikoz, J. W. F. Valle, in preparation.

%\bibitem{PDG96}
%Partcle Data Group,   Phys. Rev. D54  (1996) 1.
%(see also URL: http://pdg.lbl.gov/).


\end{thebibliography}
\end{document}